# A shift in Jupiter's equatorial haze distribution imaged with the Multi-Conjugate Adaptive Optics Demonstrator at the VLT.

Michael H. Wong[1], Franck Marchis[1,2], Enrico Marchetti[3], Paola Amico[4],
Sebastien Tordo[3], Hervé Bouy[5], Imke de Pater[1].

1: Astronomy Department, University of California, Berkeley CA 94720-3411. mikewong@astro.berkeley.edu.
2: SETI Institute, Carl Sagan Center, Mountain View CA. 3: European Southern Observatory, Garching, Germany.
4: European Southern Observatory, Paranal, Chile. 5: Instituto de Astrofísica de Canarias, La Laguna (Tenerife), Spain.

## Abstract

Jupiter was imaged during the Science Demonstration of the MCAO Demonstrator (MAD) at the European Southern Observatory's UT3 Very Large Telescope unit. Io and Europa were used as natural guide stars on either side of Jupiter, separated from each other by about 1.6 arcmin from 23:41 to 01:32 UT (2008 Aug 16/17). The corrected angular resolution was 0.090 arcsec across the entire field of view, as measured on background stars.

The observations at 2.02, 2.14, and 2.16 μm were sensitive to portions of the Jovian spectrum with strong methane absorption. The data probe the upper troposphere, which is populated with a fine (~0.5 μm) haze. Two haze sources have been proposed: lofting of fine cloud particles into the stable upper troposphere, and condensation of hydrazine produced via ammonia photochemistry. The upper tropospheric haze is enhanced over Jupiter's equatorial region.

Dramatic changes in the underlying cloud cover—part of the 2006/2007 "global upheaval"—may be associated with changes in the equatorial haze distribution now evident in the 2008 MAD images (Fig. 1). Haze reflectivity peaked at 5° N in HST/NICMOS data from 2005, but it now peaks at the equator. The observations suggest that haze variation is controlled by particle size, cloud source variation, diffusion, and horizontal transport.

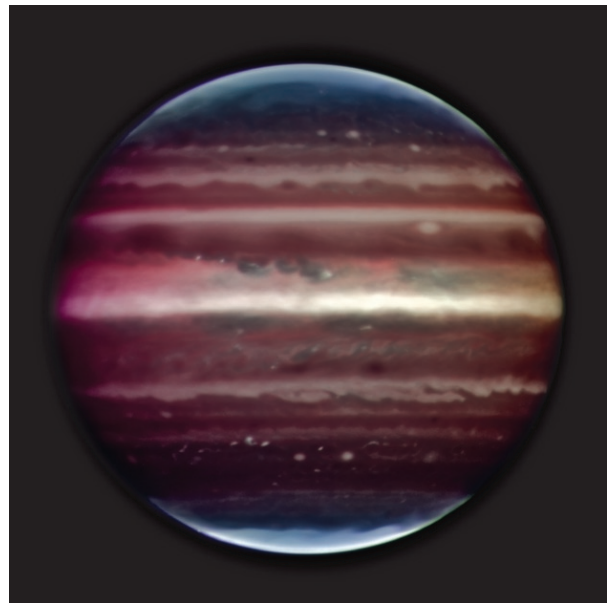

**Figure 1.** False color infrared image acquired on 2008-08-17 by the Multi-Conjugate Adaptive Optics Demonstrator (MAD) prototype instrument at the Very Large Telescope. Data in the RGB channels were acquired respectively at 0:38 UT (2.024 $^{+0.024}_{-0.054}$ μm), 0:27 UT (2.142 $^{+0.011}_{-0.009}$ μm), and 0:45 UT (2.158 $^{+0.013}_{-0.005}$ μm). To corrrect for rotation, images at 2.142 μm and 2.158 μm were re-projected onto the image at 0:38 UT, but limb darkening was not corrected, leading to coloration artifacts at Jupiter's east and west limbs. Contrast has been enhanced to show detail, but no sharpening or deconvolution has been applied.

## Observations

MAD was a prototype system providing the first wide-field AO correction for atmospheric turbulence (Marchetti et al. 2006). Two deformable mirrors conjugated at 0 and 8.5 km above the telescope provided MCAO-corrected light to a 2048 x 2048 science camera with a pixel scale of 0.028″.

The data set consists of 265 frames taken with three different filters. Galilean satellites on both sides of the planet were used to perform wavefront sensing by means of Shack-Hartmann sensors (Fig. 2). MAD achieves the best correction with three guide stars in a triangular distribution (Fig. 3), but since Galilean satellites appear colinear on the sky, only two guide stars were used (with minimal disadvantage compared to three colinear guide stars). Observations began shortly after Io emerged from eclipse (at 23:41 UT Aug 16) and ended when AO loops opened due to Europa's proximity to Jupiter (at 01:32 UT Aug 17).

Average seeing at optical wavelengths during the observations was 0.85″, and the AO correction provided a FWHM of 0.090″ across the field of view, as measured on faint background stars in many of the frames.

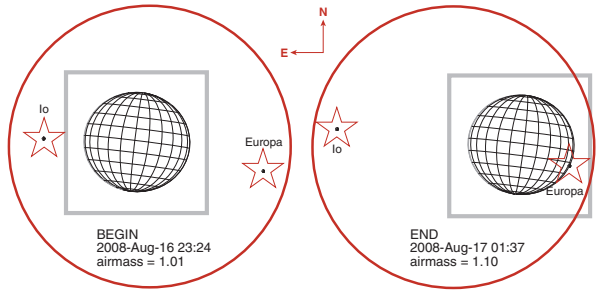

**Figure 2.** Observing geometry at beginning (end of Io eclipse) and end (onset of Europa occultation) of the experiment. The 2′ field of regard of MAD is indicated as a red circle, and the 1′ field of view of the MAD science camera is shown as a grey square. Observations were scheduled to minimize the relative motion of Europa with respect to Io. Simulated Jupiter system views were generated by the Jupiter viewer tool at the NASA PDS Rings Node.

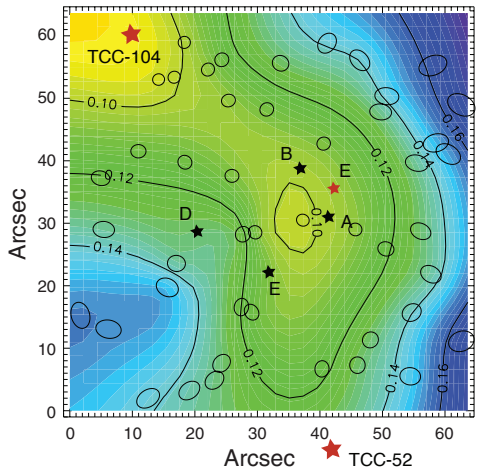

**Figure 3.** Contour map of FWHM over the field of view of the MAD science camera, measured on observations of the Trapezium cluster using three guide stars (after Bouy et al. 2008). Average seeing during the observations was 1″, at airmasses from 1.4 to 1.7. Ellipses indicate relative PSF size and shape variation across the field.

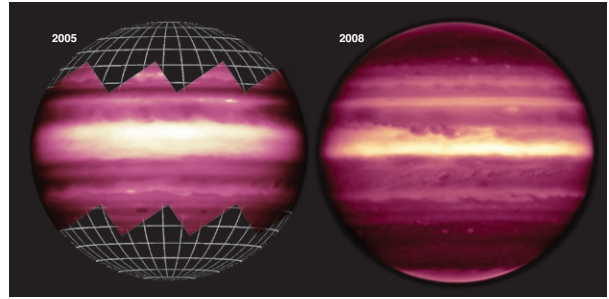

**Figure 4.** Comparison of methane-band images in 2005 and 2008, qualitatively showing changes in the equatorial haze. Left: mosaic of four NICMOS frames taken with the F212N filter as part of HST program 10161 (de Pater) on 2005-03-25 at 15:00 UT. Right: a single image from MAD at 2.02 μm taken on 2008-08-17 at 00:30 UT. In 2005, haze reflectivity was greatest north of the equator; in 2008, haze reflectivity peaked right at the equator.

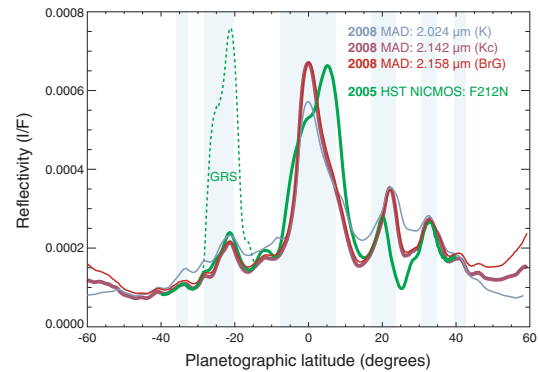

**Figure 5.** Zonally averaged haze reflectivity. The green curve shows 2005 haze reflectivity from 2.12-μm HST NICMOS data (dotted line shows increased haze reflectivity over the Great Red Spot). Other curves show 2008 haze reflectivity measured by MAD. Shaded blue areas indicate regions of anticyclonic zonal flow, measured from 2008 WFPC2 data collected for HST program 11102 (PI I. de Pater; analysis from X. Asay-Davis, personal communication). NICMOS data are displayed in standard I/F units; MAD data are scaled (but not offset) to match the NICMOS I/F at −3° and +34° because absolute calibration is not yet complete.

Data were flat-fielded and sky-subtracted using the Eclipse software (Devillard 1997), bad pixels were corrected using IRAF (Tody 1993), and deprojection into latitude-longitude space was done by matching Jupiter's limb with custom software.

## Equatorial haze changes

Figure 4 qualitatively demonstrates the dramatic change in equatorial haze distribution. The 2005 NICMOS mosaic (left) shows Jupiter's midlatitude regions and was acquired over a period of 10 minutes; the 2008 MAD image (right) shows Jupiter's entire disk in one 2-sec exposure, using the wide-area correction and large FOV of MAD. The latitude of peak haze reflectivity shifted southwards between 2005 and 2008.

To quantitatively analyze the changes, haze reflectivity was averaged over a range of longitudes and plotted as a function of latitude in Fig. 5. The change at the equator (compared to 2005) is a combination of decreased reflectivity near +5° latitude and increased reflectivity

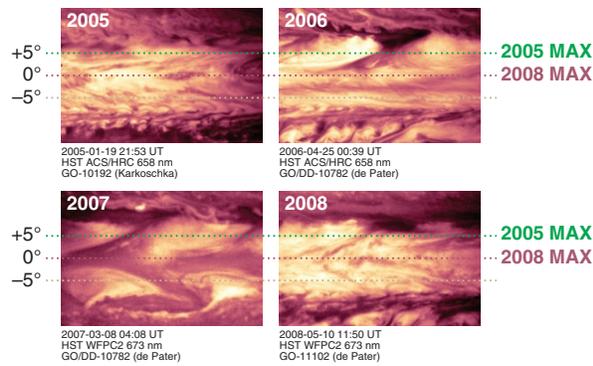

**Figure 6.** Visible wavelength HST maps spanning 45° of longitude and 30° of latitude, showing changes in underlying cloud opacity from 2005–2008. Intensity scales are linear and normalized to minimum/maximum values in each frame. Dotted lines indicate the latitude of maximum equatorial haze reflectivity in 2005 (+5°) and in 2008 (0°).

near 0°. The strengthening and northward shift of the haze band near +20° is also apparent.

Understanding the formation of the upper tropospheric haze is key to explaining the observed changes. The fine aerosols composing the tropospheric haze may be composed of condensed hydrazine (a product of ammonia photolysis at the altitude of the haze layer) mixed with smaller amounts of hydrocarbon and other photochemical products drifting down from the stratosphere (Atreya et al. 1977, 2005). Cloud particles from deeper in the troposphere may also be lofted into the haze region (West et al. 1986), where 0.5–1.0 μm particles precipitate on a timescale of 1.6 to 160 years (Rossow 1978). Due to the large energies needed to penetrate into the stably stratified haze region—energies that can be provided by moist convection—particle compositions may include $NH_4SH$ and $H_2O$ as well as $NH_3$ (Sugiyama et al. 2007). Either the photochemical or the cloud source mechanism could relate changes in haze to changes in the underlying clouds during Jupiter's 2006/2007 global upheaval.

## The 2006/2007 upheaval

Visible-light reflectivity maps in Fig. 6 show the two main changes in equatorial cloud patterns during the upheaval: overall cloud reflectivity decreased, and south equatorial disturbances (SEDs) appeared (fan-like plumes near −5° latitude in the 2007 frame). Little is known about SEDs, but their methane-band reflectivities indicate that they do not reach the same high altitudes as the large convective plumes which erupted in 2007 at +23° latitude (Sánchez-Lavega et al. 2008). The injection of haze at high altitudes by the plumes at +23° is the probable cause of the haze enhancement there (Fig. 5).

Could the SEDs have nonetheless injected enough material to account for the change in the equatorial haze? If so, meridional transport would need to be invoked because the haze maximum appears 5° to the north of the SEDs. Due to the decreased cloud reflectivity at 0° during the upheaval (Fig. 6), the direct cloud source should have led to a decrease in haze at the equator, rather than the observed increase. A diffusive transport model (Wong 2007) showed that a complete cutoff of the cloud source would decrease the column density of 0.5-μm particles by 40% in one year, a change of similar magnitude to the observed drop at +5° between 2005 and 2008.

An increase in condensed hydrazine haze would naturally result from a cooling of the radiatively-warmed upper troposphere. But would decreased equatorial cloud opacity produce a net warming due to increased upwelling infrared radiation, or a net cooling due to less solar radiation reflected back from deeper clouds? A quantitative study of the radiative balance is needed to answer this question, as well as high time-resolution observations (several observations per year) of upper tropospheric temperatures.

## Acknowledgements
We thank Jorge Melnick (VLT Program Scientist) for granting VLT access for this project. Funding for UC Berkeley collaborators was provided by the NSF Center for Adaptive Optics and the Space Telescope Science Institute (STScI). This work is based in part on observations made with the NASA/ESA Hubble Space Telescope, obtained at STScI (operated by the Association of Universities for Research in Astronomy, Inc., under NASA contract NAS 5-26555). These observations are associated with programs 10161, 10192, 10782, and 11102.